\documentclass[aps,pre,epsfigure,twocolumn,superscriptaddress]{revtex4-1}
\usepackage[colorlinks=true,linkcolor=blue,urlcolor=blue,citecolor=blue,pdfusetitle]{hyperref}
\usepackage[utf8]{inputenc}
\usepackage[english]{babel}
\usepackage{amsmath}
\usepackage[caption = false]{subfig}
\usepackage{graphicx,epstopdf}
\usepackage{blindtext}
\usepackage[table,xcdraw]{xcolor}
\usepackage{lipsum}
\usepackage{amsfonts}
\usepackage{bbm}
\usepackage{amssymb}
\usepackage{enumerate}
\usepackage{color}
\usepackage{latexsym}
\usepackage{physics}
\usepackage{times,txfonts}

\newcommand{\Ccal}{\mathcal{C}}
\newcommand{\Dcal}{\mathcal{D}}
\newcommand{\Ecal}{\mathcal{E}}

\newcommand{\Hcal}{\mathcal{H}}

\newcommand{\Ncal}{\mathcal{N}}

\newcommand{\Rcal}{\mathcal{R}}

\newcommand{\1}{\mathbbm{1}}

\newcommand{\interpro}[2]{\langle #1 | #2 \rangle}

\begin{document}

\title{Quantum advantage of two-level batteries in self-discharging process}

\author{Alan C. Santos}
\email{ac\_santos@df.ufscar.br}
\affiliation{Departamento de Física, Universidade Federal de São Carlos, Rodovia Washington Luís, km 235 - SP-310, 13565-905 São Carlos, SP, Brazil}

\date{\today}

\begin{abstract}
Devices that use quantum advantages for storing energy in the degree of freedom of quantum systems have drawn attention due to their properties of working as quantum batteries (QBs). However, one can identify a number of problems that need to be adequately solved before a real manufacturing process of these devices. In particular, it is important paying attention to the ability of quantum batteries in storing energy when no consumption center is connected to them. In this paper, by considering quantum batteries disconnected from external charging fields and consumption center, we study the dissipative effects that lead to charge leakage to the surrounding environment. We identify this phenomena as a \textit{self-discharging} of QBs, in analogy to the inherent decay of the stored charge of conventional classical batteries in a open-circuit configuration. The performance of QBs concerning the classical counterpart is highlighted for single- and multi-cell quantum batteries.
\end{abstract}

\maketitle

\section{Introduction}

Over the last few years, one has been paid some attention to the development of quantum devices able to store energy to be used for later processing. The result after years of studies is a number of works discussing the performance of quantum batteries (QBs) concerning their charging power~\cite{Alicki:13,Ferraro:18,Andolina:18,Crescente:20,Rossini:20,PRL_Andolina,Santos:19-a,Santos:20c,Kamin:20-2,Kamin:20-1,James:20} and work extraction~\cite{PRL2013Huber,CampbellBatteries,Fusco:16,Gianluca:17,Alexia:20}. One can highlight as a significant advance to this field the work developed by Allahverdyan, Balian and Nieuwenhuizen, who addressed the question of how much energy can be extracted from quantum system by unitary operations~\cite{Allahverdyan:04}. Based on quantum thermodynamics theory, they showed that the extractable maximum energy is given by
\begin{align}
\Ecal = \sum\nolimits_{i=1}^{\Dcal} \sum\nolimits_{n=1}^{\Dcal} \varrho_{n}\epsilon_{i} \left( |\interpro{\varrho_{n}}{\epsilon_{i}}|^2 - \delta_{ni} \right) . 
\label{Ergo2}
\end{align}
also known as \textit{ergotropy}, with $\Dcal$ being the dimension of the Hilbert space and we write $H_{0}\! =\!\sum\nolimits_{n=1}^{\Dcal} \epsilon_{n} \ket{\epsilon_{n}}\bra{\epsilon_{n}}$ and $\rho\! =\!\sum\nolimits_{n=1}^{\Dcal} \varrho_{n} \ket{\varrho_{n}}\bra{\varrho_{n}}$, so that $\varrho_{1}\!\geq\!\varrho_{2}\!\geq\!\cdots\!\geq\! \varrho_{\Dcal}$ and $\epsilon_{1}\!\leq\!\epsilon_{2}\!\leq\!\cdots\!\leq\! \epsilon_{\Dcal}$.

However, it is important to keep in mind we are still far from a real development of practical QBs due to a number of different reasons. In fact, if we understand that the manufacturing process of these devices is not only justified by its high charging power, we raise a number of question that have not yet been adequately addressed. As one of these questions, we highlight here the phenomenon known as \textit{self-discharging} (SD) of batteries~\cite{Conway:97,Ricketts:00,Kowal:11,Lei-Zhang:18,Zhang:20}. This process leads to the loss of charge due to inherent characteristic of the system used as working fluid for storing energy and it happens regardless whether the battery is connected to some consumption hub. Although we have a number of proposals of two-level QBs in different systems, such as spin systems~\cite{Le:18}, 
quantum cavities~\cite{Binder:15,Fusco:16,Zhang:18,Ferraro:18}, among others~\cite{Andolina:18,Andolina:19,Bai:20,Long:03}, the SD mechanism for this kind of QB is yet an open question.

Based on studies of SD in commercial batteries, in this paper we introduce a strategy to study SD processes in QBs. The key point is defining how to put a QB as an open-circuit, which is intuitively done by the absence of external charging or extracting energy fields, and interaction with auxiliary systems used as consumption hubs. We apply our approach to single- and multi-cell QBs, where we show that by using quantum properties of the system (more specifically, quantum state superposition) the performance of the storing energy device is enhanced concerning the classical counterpart  (no quantum superposition).

\begin{figure}
	\centering
	\includegraphics[scale=0.22]{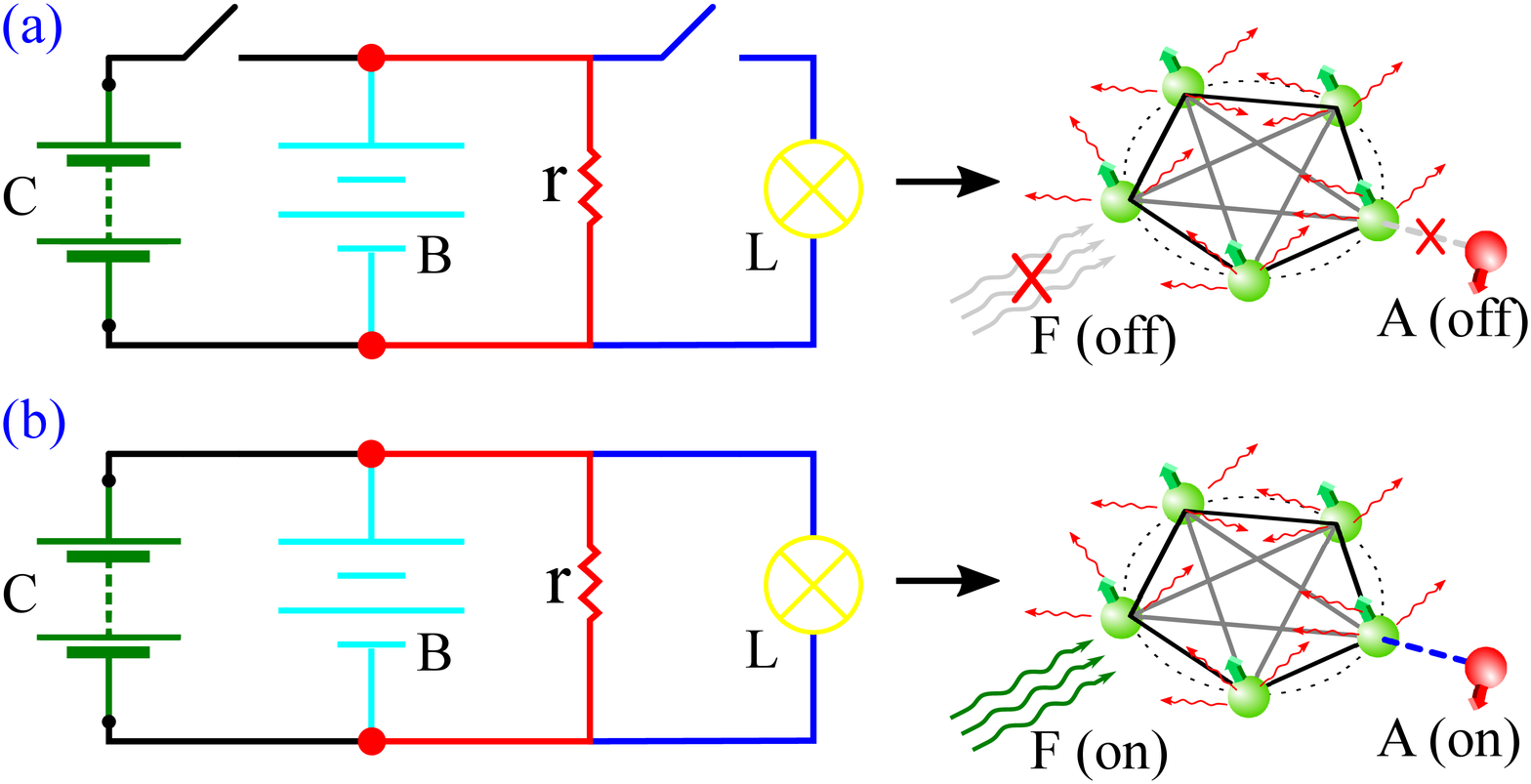}\label{CloseFig}
	\caption{Sketch of {\color{blue}(a)} an classical open circuit consists of a charger C, the battery B, an internal resistance r and an external circuit L (a lamp), where the resistance r is considered to describe the ohmic self-discharging of a classical battery~\cite{Kowal:11}. The arrow shows the equivalent of the classical circuit in a quantum approach. An external field F is used as charger and an auxiliary qubit A is used as consumption center (the ``lamp"). {\color{blue}(b)} Classical open-circuit configuration used to study SD processes and its quantum counterpart as we are proposing here to do similar studies of SD in QBs.}\label{FigScheme}
\end{figure}

\section{Single-cell QBs}

The study of SD in classical batteries is done in the following way: after charging the battery or capacitor, we let the system evolves in a open-circuit configuration, where inevitably some amount of energy is lost, leading then to the SD process. As illustrated in Fig.~\ref{FigScheme}, a similar approach can also be applied to QBs. The dynamics of a QB can be written as
\begin{align}
\dot{\rho}(t) = \Hcal[\rho(t)] + \Rcal[\rho(t)] ,
\end{align}
where $\Hcal[\rho(t)]$ describes the unitary dynamics of the system and/or coherent interaction with consumption hub and $\Rcal[\rho(t)]$ encodes all loss processes due to the coupling with an environment. In particular, let us also suppose that 
\begin{align}
\Hcal[\rho(t)] = \frac{1}{i\hbar} [H_{0} + H_{\text{int}} + H_{\text{ext}}(t),\rho(t)] ,
\end{align}
with $H_{\text{int}}$ is a Hamiltonian used to deal with internal interactions between the cells of the battery $H_{\text{ext}}(t)$ being an arbitrary field used to inject or extract energy from the QB (external field). Then, we say that the battery is in a open-circuit configuration when $H_{\text{ext}}(t)\!=\!0$, situation in which one can adequately study the SD of QBs using a similar strategy as done for classical batteries.

As a first case, let us consider the most elementary case of a QB constituted by a single two-level system with internal Hamiltonian $H_{0}\!=\!\hbar \omega \sigma^{+}\sigma^{-}$, $\sigma^{-}\!=\!\sigma^{+\dagger}\!=\!\ket{g}\bra{e}$, driven by the master equation
\begin{align}
\dot{\rho}(t) = \frac{1}{i\hbar} [H_{0},\rho(t)] + \Rcal_{\text{rel}}[\rho(t)] ,
\end{align}
which describes the relaxation processes leading to energy dissipation, with
\begin{align}
\Rcal_{\text{rel}}[\rho(t)] &= \frac{\Gamma}{2} \left( 2\sigma^{-} \rho(t) \sigma^{+} - \{\sigma^{+}\sigma^{-},\rho(t)\} \right) , \label{Eq-RelSingle}
\end{align}
being $\sigma^{z}\!=\!\ket{e}\bra{e}-\ket{g}\bra{g}$. In particular, we assume this process in our study because it is the most common decoherence processes in a large of physical systems, e.g. in nuclear spin systems~\cite{Peterson:18,Sarthour:Book,Hong:17}, cavity quantum electrodynamics~\cite{Garraway:97} and superconducting qubits~\cite{Peterer:15,Wen:19,lin:19}. Then, by writing the density matrix in the QB basis $\ket{e}$ and $\ket{g}$ as
\begin{align}
	\dot{\rho}(t) = \varrho_{\text{e}}(t) \ket{e}\bra{e} + \varrho_{g}(t) \ket{g}\bra{g} + \left( \varrho_{\text{eg}}(t) \ket{e}\bra{g} + \text{h.c.}\right) ,
\end{align}
the solution of the dynamics can be found as
\begin{align}
	\varrho_{\text{e}}(t) &= \varrho_{\text{e}}(0) e^{-\Gamma t} ~,~~ \varrho_{\text{eg}}(t) = \varrho_{\text{eg}}(0) e^{-i\omega t} e^{-2 \Gamma t} .
\end{align}

From this general result for the dynamics considered here, we can study different situations. In particular, we are interested in the decay performance when we use quantum resources (coherence) to store energy in the QB.

\subsection{Discharging from full charge state}

As a first discussion, let us assume the case of a battery initially in a full charged state $\varrho_{\text{eg}}(0)\!=\!0$ and $\varrho_{\text{e}}(0)\!=\!1$. In this case, the available energy as given by the ergotropy $\Ecal(t)$ and the internal energy $U(t)$ read (with $\Ecal_{\text{max}}\!=\!\hbar \omega$)
\begin{align}
\Ecal(t) = \Theta(\tau_{\text{c}} - t) \left( 2 e^{-\frac{t\Gamma}{2}} -1 \right) \Ecal_{\text{max}} ,~~ U(t) = \Ecal_{\text{max}}e^{-\frac{t\Gamma}{2}} ,
\end{align}
where $\Theta(x)$ is the Heaviside theta ($0$ if $x\!<\!0$ and $1$ if $x\!\geq\!0$) and $\tau_{\text{c}}$ is the population crossing time. This crossing is inevitable due to the decay where the instantaneous population in excited $\varrho_{\text{e}}(t)$ decreases from $1$ to $0$ and the population in ground state $\varrho_{\text{g}}(t)$ goes through the contrary direction. Then, $\varrho_{\text{e}}(t)\! >\!\varrho_{\text{g}}(t)$ for $t\!<\!\tau_{\text{c}}$, $\varrho_{\text{e}}(t)\!=\!\varrho_{\text{g}}(t)$ for $t\!=\!\tau_{\text{c}}$, and $\varrho_{\text{e}}(t)\! < \!\varrho_{\text{g}}(t)$ for $t\!>\!\tau_{\text{c}}$, consequently, the basis ordering requested by the Eq.~\eqref{Ergo2} changes in time during the dynamics. From a simple calculation by imposing $\varrho_{\text{e}}(t)\!=\!\varrho_{\text{g}}(t)$ we can show that $\tau_{\text{c}}\!=\!2\ln(2)/\Gamma$.

The decay of both quantities $\Ecal(t)$ and $U(t)$ is well-described by the monotonic decreasing exponential $e^{-t/\tau_{\text{d}}}$, with $\tau_{\text{d}}\!=\!2/\Gamma$ being a characteristic decay time scale. This kind of decay process is similar to the ohmic behavior of SD for conventional classical batteries~\cite{Conway:97,Ricketts:00}, where we also get a description in terms of a single decay time scale. Then we understand this result as intuitively expected due to the absence of any quantumness in the battery during the process (quantum state coherence, for example). It is also worth highlighting that the ergotropy sudden death arising to the population ordering during the system evolution. With this example we argue that, although we have some quantum advantage in the charging process of the a two-level system~\cite{Andolina:19}, the fact of a quantum system to be charged in the full charge state cannot be taken as quantum concerning SD processes, since its behavior is equivalent to conventional classical batteries.

\subsection{Exploring quantum advantage}

\begin{figure}[t!]
	\centering
	\includegraphics[scale=0.37]{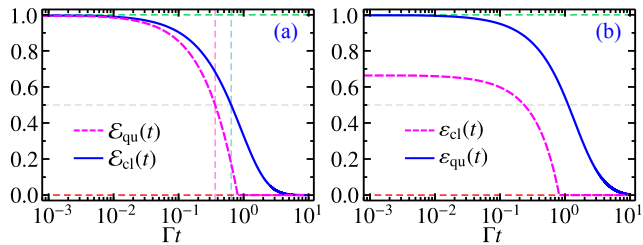}
	\caption{{\color{blue}(a)} Ergotropy (as a multiple of $\Ecal_{1/2}$) for the initial classical and quantum states. Vertical dashed lines denotes the respective half-life time for classical and quantum states given, respectively, by $\tau_{1/2}^{\text{cl}}\Gamma\!=\!2\ln(6/5)\!\approx\!0.36$ and $\tau_{1/2}^{\text{qu}}\Gamma\!=\!2\ln[2 (1+\sqrt{6})/5 ]\!\approx\!0.65$. {\color{blue}(b)}~Efficiency coefficient $\varepsilon(t)$ for the same process.}\label{Fig1}
\end{figure}

Now, let us explore quantum coherence of a two-level QB to study the same process. As an immediate consequence of considering quantum coherence, we remark the impossibility of starting the system in a full charged state. Hence, in order to have a fair comparison between classical and quantum performance we will assume that the system starts with ergotropy $\Ecal(0)\!=\!\Ecal_{1/2}\!=\!\Ecal_{\text{max}}/2$. Under this choice, we can compare the gain of an state with maximal coherence $\ket{\psi}\!=\!(\ket{e}+\ket{g})/\sqrt{2}$ regarding its classical counterpart given by the mixed classical state $\rho_{\text{cl}}\!=\!(3/4)\ket{e}\bra{e}+(1/4)\ket{g}\bra{g}$. It is straightforward conclude that the ergotropy for both states is $\Ecal_{1/2}$. Then, the ergotropy for the classical and quantum version, respectively, give
\begin{subequations}
\begin{align}
\Ecal_{\text{cl}}(t) &= \Ecal_{1/2}\Theta\left(\tau_{\text{c}}^{(1/2)} - t\right) \left( 3 e^{-\frac{t\Gamma}{2}} - 2 \right) \hbar\omega , \\
\Ecal_{\text{qu}}(t) &=  \Ecal_{1/2}e^{-t\Gamma/2} \left( 1 - e^{t\Gamma/2} \sqrt{ 2 - 2e^{t\Gamma/2} + e^{-t\Gamma} }\right)  ,
\end{align}
\end{subequations}
where $\tau_{\text{c}}^{(1/2)}\!=\!2\ln(3/2)/\Gamma$. Then, it is immediate to conclude some substantial difference between the two initial states. Different from the classical case, the existence of quantum coherence leads us to a SD phenomena that cannot be explained by an ohmic process, a characteristic of commercial classical supercapacitors~\cite{Kowal:11,Lei-Zhang:18,Zhang:20}. In the Fig.~\ref{Fig1}{\color{blue}a} we show the time evolution of the above quantities, highlighting the half-life time $\tau_{1/2}$ (concerning the initial amount of stored energy) for each kind of battery. Then, one can see the advantage of storing energy in quantum information concerning classical one. 

For completeness, we also consider the portion of energy stored in the system as ergotropy from quantity $\varepsilon(t) = \Ecal(t)/U(t)$, being $U(t)\!=\!\tr(\rho(t)H_{0})$ the internal energy of the system. We understand the parameter $\varepsilon(t)$ in the following way: the energy conservation law says that we need to spend a total amount of energy $U(t)$ to store an total $\Ecal(t)$ of ergotropy in the QB, so that some portion of energy cannot be extracted from the QB by unitary operations. Thus, it gives us a way to quantify the percentage of $U(t)$ that is wasted in the process. In fact, in case $\varepsilon(t)\!=\!1$ we have $U(t)\!=\!\Ecal(t)$, while $\varepsilon(t)\!<\!1$ means $U(t)\!>\!\Ecal(t)$ and hence there is an amount of `dead energy' that cannot be later extracted by unitary operations, being $\varepsilon(t)\!=\!0$ the worst scenario. The result is shown in Fig.~\ref{Fig1}{\color{blue}b}. In this scenario, the advantage of QBs becomes even more evident, in which we can design a battery that simultaneously presents a decay slower than its classical counterpart, and spends less energetic resource to store useful work.

\subsection{Quantum supremacy of QBs in self-discharging}	
	
With recent progress in quantum control, the processing power of quantum computers has reached a frontier close to \textit{quantum supremacy}~\cite{Arute:19,Preskill:12}. In a few words, quantum supremacy happens when a quantum device implements a given task that a classical one is unable to perform or needs a impracticable time to do it. Then, it leads us to the question whether QBs provide absolute advantage concerning classical ones. In this sense, it is possible to show that QBs do not present quantum supremacy regarding self-discharging processes. We mean, there is at least one kind of channel in which storing energy in coherence is not a good choice concerning the classical counterpart. In fact, let us consider the same initial states as previously for a single-cell QB case. In a scenario in which the system evolves under the phase damping channel~\cite{Nielsen:Book}, the evolved state for the classical state reads $\rho_{\text{cl}}(t)\!=\!\rho_{\text{cl}}$, since the phase damping channel does not affect the battery, and the ergotropy is not lost. On the other hand, the same channel acting on the quantum state gives us~\cite{Nielsen:Book}
\begin{align}
\rho_{\text{qu}}(t) = \frac{1}{2} \left(\ket{e}\bra{e} + \ket{g}\bra{g}\right) + \frac{e^{-\gamma t}}{2}  \left(\ket{e}\bra{g} + \ket{g}\bra{e}\right) .
\end{align}
where $\gamma$ is the damping rate. Therefore, in the asymptotic regime $t\!\gg\!\gamma$ we find $\rho_{\text{qu}}(t)\!\approx\!(1/2)\left[\ket{e}\bra{e} + \ket{g}\bra{g}\right]$, a state with zero ergotropy. In conclusion, while classical state is able to keep ergotropy for an infinity time for this kind of dissipation, the ergotropy stored in quantum states is quite volatile. It is worth highlighting that phase damping is a purely quantum mechanical effect which promotes the loss of information (quantum superposition) without loss of energy. Then, since ergotropy of quantum states is stored as quantum information, it is drastically affected by this kind of decoherence.

\section{N-cell QBs}

In general we also are interested in exploring quantum effects in multi-cell QBs as, for example, entanglement or coherence. By considering a linear chain of $N$ cells, we let the system evolving under a decay process described by the master equation considered as
\begin{align}
\dot{\rho}(t) = \frac{1}{i\hbar} [H_{0}(N) + H_{\text{int}}^{\text{lin}},\rho(t)] + \Rcal_{\text{rel}}[\rho(t)] + \Rcal_{\text{icl}}[\rho(t)] , \label{Eq-Shro-Ncell}
\end{align}
where $H_{0}(N)\!=\!\sum_{n=1}^{N}\hbar \omega \sigma_{n}^{+}\sigma_{n}^{-}$, $ H_{\text{int}}$ is the interaction Hamiltonian between cells of the battery given as $H_{\text{int}}^{\text{lin}}\!=\! \sum_{n} J \hbar \sigma^{-}_{n}\sigma^{+}_{n+1}$, $\Rcal_{\text{rel}}[\rho(t)]$ being independent relaxation, as in Eq.~\eqref{Eq-RelSingle} for each cell, and the second term reads
\begin{align}
\Rcal_{\text{icl}}[\rho(t)] &= \sum_{n = 1}^{N} \frac{\Gamma_{\text{icl}}}{2} \left( 2\sigma^{-}_{n} \rho(t) \sigma^{+}_{n+1} - \{\sigma^{+}_{n+1}\sigma^{-}_{n},\rho(t)\} \right) \nonumber \\
&+\sum_{n = 1}^{N} \frac{\Gamma_{\text{icl}}}{2} \left( 2\sigma^{+}_{n} \rho(t) \sigma^{-}_{n+1} - \{\sigma^{-}_{n+1}\sigma^{+}_{n},\rho(t)\} \right) ,
\end{align}
that describes the intracell leakage, in which a given cell $n$ of the battery would emit an excitation and then this excitation is temporarily stored in its nearest-neighbor cell. 
	
The master equation above describes a $N$-cell QB where the cells interact each other. The interaction Hamiltonian $H_{\text{int}}^{\text{lin}}$ describes the interaction of a linear chain in different contexts. For example, it describes the capacitive interaction between two superconducting qubits~\cite{majer2007,Xu:20}, or the dipole-dipole interaction in atomic chain~\cite{Lehmberg:70-I,Lehmberg:70-II,Cidrim:20,Masson:20}. The term $\Rcal_{\text{icl}}[\rho(t)]$ in the master equation describes an collective decay of the system, in which the $n$-th cell has its decay dependent on the other cells of the battery. The master equation considered here describes a system of $N$ quantum emitters (two-level atoms or artificial atoms) interacting with a continuum of photonic modes, where the photonic degrees of freedom are adequately eliminated by a Born-Markov approximation. Then, after this process we find a master equation for $N$ interacting emitters mediated by virtual photons~\cite{Lehmberg:70-I,Lehmberg:70-II,Cidrim:20,Masson:20,majer2007,Xu:20}. In our analysis we consider a dissipative channel $\Rcal^{\text{lin}}_{\text{icl}}[\rho(t)]$ in which the nearest-neighbor interaction is the most relevant contribution. It is worth mentioning that in case where the mediated interaction is negligible, one gets $H_{\text{int}}^{\text{lin}}\!=\!0$ and $\Rcal_{\text{icl}}[\rho(t)]\!=\!0$, so that the system dynamics is given by a set of independent cells as studied in the previous section.

\begin{figure}[t!]
	\centering
	\includegraphics[scale=0.37]{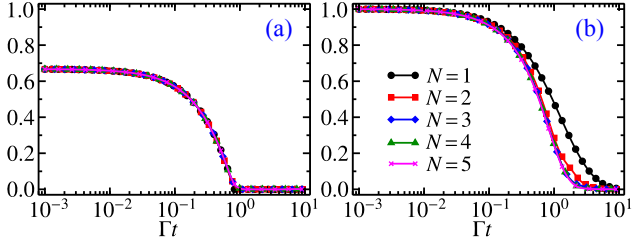}
	\caption{Efficiency coefficient $\varepsilon(t)$ for {\color{blue}(a)} classical and {\color{blue}(b)} quantum states for some values of $N$. Here we consider $J\!=\!2 \Gamma$ and $\Gamma_{\text{icl}}\!=\!\Gamma/2$.}
	\label{Fig2}
\end{figure}

It is worth mentioning that the internal energy $U$ we are interested does not take into account the Hamiltonian $H_{\text{int}}^{\text{lin}}$, since the existence of interactions will promote positive/negative contributions in $U$ and it eventually lead to an unfair comparison when we consider different kind of geometry beyond the linear one (as we shall consider soon). For this reason, only contributions of the QB energy basis is taken into account, we mean, $U\!=\!\tr(H_{0}(N)\rho)$. We analyze the dynamics in the same scenario as before by taking the classical state of $N$ cells as $\rho_{\text{cl}}(N)\!=\!\otimes_{n=1}^{N} \rho_{\text{cl}}$ and the quantum state as $\ket{\psi(N)}\!=\!\otimes_{n=1}^{N}\ket{\psi}$. Because each cell starts in a configuration with ergotropy $\Ecal_{1/2}$, then the initial amount of ergotropy in the QB as function of $N$ is $\Ecal_{1/2}(N)\!=\!N/2$. In Fig.~\ref{Fig2}{\color{blue}a} we show the efficiency parameter $\varepsilon(t)$ for the classical state, normalized by the initial amount of ergotropy in each case, and the Fig.~\ref{Fig2}{\color{blue}b} shows the same quantity for the quantum version. 

As in the single-qubit case, here we can see enhancement for $N$-cells quantum battery (at least up to $N\!=\!5$) concerning the discharging time. In addition, by taking into account that the initial amount of efficiency in storing energy as ergotropy, as computed by the parameter $\varepsilon(0)$, it shows a clear quantum advantage scaling with $N$. In fact, the value $\varepsilon_{\text{qu}}(0)\!=\!1$ means high efficiency of the quantum version for any $N$. On the other hand, the quantity $\varepsilon_{\text{cl}}(0)\!\approx\!0.667$ says we have a total of `dead energy' given by $\sim33.3\%$ per cell, hence it is easy to see that the total amount of worst energy scales as $\sim 0.333 \hbar \omega N$. It is worth mentioning that due to the initial state considered here, no quantum advantage is obtained from entanglement, and that makes our proposal useful due to the very simple state we use as resource to store energy.

\subsection{Exploring network effects}

It is a reasonable assumption to think that the kind of interactions between the cells of an QB would develop some role in its charging process. For this reason, it is worth analyzing how the system geometry affects the SD. In this scenario, we write the interaction Hamiltonian in a general way as $H_{\text{int}}\!=\! \sum_{\{n,m\}} J \hbar \sigma^{-}_{n}\sigma^{+}_{m}$, in which the sum $\sum_{\{n,m\}}$ is done over all direct interactions between the QB cells. In addition, we assume that connection between each pair of cells induces a collective decay channel each other, so that the master equation is modified as
\begin{align}
\Rcal_{\text{icl}}[\rho(t)] &= \sum_{\{n,m\}} \frac{\Gamma_{\text{icl}}}{2} \left( 2\sigma^{-}_{n} \rho(t) \sigma^{+}_{m} - \{\sigma^{+}_{m}\sigma^{-}_{n},\rho(t)\} \right) .
\end{align}

\begin{figure}[t!]
	\centering
	\includegraphics[scale=0.37]{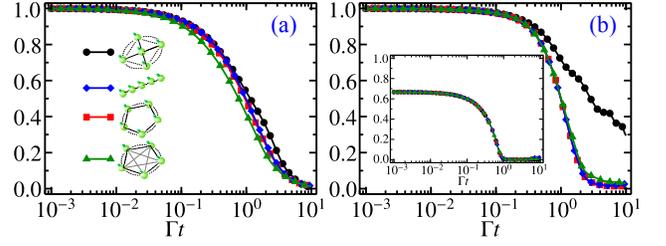}
	\caption{{\color{blue}(a)} Internal energy as a multiple of the initial internal energy $U_{0}\!=\!\bra{\psi(N)}H_{0}(N)\ket{\psi(N)}\!=\!N\hbar\omega/2$. {\color{blue}(b)} Efficiency coefficient $\varepsilon(t)$ for the (main) quantum and (inset) classical states. Here we assume $N\!=\!5$, $J\!=\!2 \Gamma$ and $\Gamma_{\text{icl}}\!=\!\Gamma/2$.}
	\label{Fig3}
\end{figure}

This kind of dynamics is usually found in light mediated interaction in atomic systems~\cite{Lehmberg:70-I,Lehmberg:70-II,Cidrim:20} and superconducting qubits~\cite{Wen:19,lin:19}, for example. To illustrate our results, we consider four kind of geometries for a five-cell QB. They are named here as central, linear, circular and symmetric network, as shown in the legends of the Fig.~\ref{Fig3} (from top to bottom, respectively).

It is intuitive to imagine that the more intracell decay, the slower the energy leakage from the system into the surrounding environment. 
On the other hand, consider the result shown in Fig.~\ref{Fig3}{\color{blue}a}, in which we compute the instantaneous internal energy $U(t)\!=\!\tr(\rho(t)H_{0}(N))$ for the quantum state, being $\rho(t)$ the the solution of the Eq.~\eqref{Eq-Shro-Ncell}. This behavior suggests a complementary discussion, being the topology most relevant to the collective energy decay than the number of intracell channels. As example, we can mention the linear and central geometries, in which they have a same number of connections but they present different property of storing internal energy (excitation). In addition, and most important here, the efficiency in storing energy as ergotropy is strongly affected by the geometry. As a highlight, by using a central geometry we guarantee more efficiency than in the other cases, as it is shown in Fig.~\ref{Fig3}{\color{blue}b}. In addition to keep energy for more time, the capacity of storing ergotropy is also enhanced relatively to the other geometries. In conclusion, we can see that the connectivity between each cell of the QB can be used as resource to design QBs with different SD characteristics. If we use classical states to used store ergotropy the connectivity does not matter, as we can see in Fig.~\ref{Fig3}{\color{blue}b} (inset).

In order to give more details about the quantumness of the battery, we also consider the dynamics of the entanglement and coherence of the system during the dynamics. As measure of entanglement, we use the pairwise concurrence between two cells and we take into account an average over all connections of the battery. Mathematically this reads
\begin{align}
\Ccal(t) = (1/\Ncal_{\{n,m\}})\sum\nolimits_{\{n,m\}} \Ccal_{nm}(t) ,
\end{align}
where $\Ncal_{\{n,m\}}$ denotes the number of total connections in the battery, for example, one has $\Ncal_{\{n,m\}}\!=\!4$ and $\Ncal_{\{n,m\}}\!=\!5$ for the central and circular networks, respectively, in a five-cell QB. To end, the pairwise concurrence $\Ccal_{nm}(t)$ is obtained by taking the reduced density matrix of the $n$-th and $m$-th cells. This quantity is computed from definition of concurrence by Hill-Wootters that reads $\Ccal(\hat\rho)\!=\!\max \{ 0, \lambda_{1} - \lambda_{2} - \lambda_{3} - \lambda_{4}\}$,
where $\lambda_{1},\cdots,\lambda_{4}$ are the eigenvalues in decreasing order of the matrix $\hat R\!=\! ( \hat\rho^{1/2}\hat{\tilde{\rho}}\hat\rho^{1/2} )^{1/2}$, where $\hat{\tilde{\rho}}\!=\!(\hat\sigma_{y}\otimes \hat\sigma_{y}) \hat\rho^{*} (\hat\sigma_{y}\otimes \hat\sigma_{y})$, with $\hat\rho^{*}$ being the complex conjugate of $\hat\rho$ written in the highly entangled Bell basis~\cite{Hill:97}. On the other hand, the coherence is computed from its conventional definition based on the $l_1$-norm normalized coherence~\cite{Baumgratz:14,Streltsov:17} defined as $C(t)\!=\!(1/C_{\text{max}})\sum_{i,j\neq i} |\rho_{ij}(t)|$, with $\rho_{ij}(t)$ being the matrix elements of the system state. This measure of coherence is enough for our discussion, since our reference basis is well defined by the empty and charged states of the battery~\cite{Kamin:20-2}.

\begin{figure}[t!]
	\centering
	\includegraphics[scale=0.37]{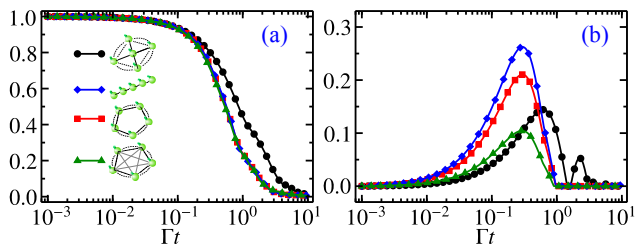}\label{Fig4-Cohe}
	\caption{{\color{blue}(a)} Coherence and {\color{blue}(b)} entanglement dynamics for the system assumed in Fig~\ref{Fig3}. Here $N\!=\!5$, $J\!=\!2 \Gamma$ and $\Gamma_{\text{icl}}\!=\!\Gamma/2$.}
	\label{Fig4}
\end{figure}

We identify some amount of coherence and entanglement that arise during the decay dynamics starting from the classical state, but they are very small quantities (of order of $10^{-2}$) and we neglected them. On the other hand, these quantities are very significant for the case where initial state of the QB has initial coherence, therefore we show these quantities in Fig.~\ref{Fig4}. The behavior for coherence and entanglement leads us to conclude that the resistance of a QB against decay process is associated with the amount of coherence in the system. In fact, while the entanglement can increase during the dynamics due to collective decay of the system, the efficiency follows an equivalent decreasing process as followed by the coherence. Moreover, to the best of our knowledgment, it is not trivial to conclude whether entanglement develops some role in the kind of process we are considering in this manuscript. We mean, from the dynamics for the central geometry we can notice that the collective decay induces both entanglement births and death. However, the efficiency cannot be perfectly explained by the entanglement behavior. This result reinforces a recent discussion raised in Refs.~\cite{Kamin:20-2,Le:18}, where it has been provided some evidences that entanglement is not the main resource for QBs. 

\section{Conclusion}

In this paper we study the SD process of two-level QBs, in which the dissipative processes lead to energy loss from the system to its surrounding environment. The quantum advantage is firstly explored in a single-cell QB, where the usage of quantum coherence to store energy leads to a slower decay concerning the process in which energy is initially stored in classical states. When we increase the system, both coherence and geometry of the intracell connections seem to be good sources to design robust QB against these undesired effects that discharges our QB, even whether it is not coupled to some consumption hub. 

It is important mentioning that this paper shows a scenario in which storing ergotropy as coherence provides an enhanced performance concerning the case where ergotropy is stored as populations. Therefore, our results suggest an advantage of using quantum properties of the system instead classical ones. No quantum supremacy of QBs is supported by this study. The study of general dissipative channels in self-discharging of quantum batteries is a good topic for future research. The results presented here open perspectives for new advances in QBs, since keeping energy stored in QB for long times after the charging process is a key task for the development of realistic energy storing quantum devices.

\begin{acknowledgments}
This research is supported by S\~ao Paulo Research Foundation (FAPESP) (Grant No 2019/22685-1).
\end{acknowledgments}




%

\end{document}